# PSEUDOSCALAR-PHOTON INTERACTIONS, AXIONS, NON-MINIMAL EXTENSIONS, AND THEIR EMPIRICAL CONSTRAINTS FROM OBSERVATIONS[1]

WEI-TOU NI,[*] A. B. BALAKIN,[†] HSIEN-HAO MEI[*]

[*]*Center for Gravitation and Cosmology, Department of Physics,
National Tsing Hua University, Hsinchu, Taiwan, 30013 Republic of China*
*weitou@gmail.com, mei@phys.nthu.edu.tw*
[†]*Kazan State University, Kremlevskaya street 18, 420008, Kazan, Russia*
*Alexander.Balakin@ksu.ru*

Pseudoscalar-photon interactions were proposed in the study of the relations among equivalence principles. The interaction of pseudoscalar axion with gluons was proposed as a way to solve the strong CP problem. Subsequent proposal of axion as a dark matter candidate has been a focus of search. Motivation from superstring theories add to its importance. After a brief introduction and historical review, we present (i) the current status of our optical experiment using high-finesse Fabry-Perot resonant cavity – Q & A experiment – to detect pseudoscalar-photon interactions, (ii) the constraints on pseudoscalar-photon interactions from astrophysical and cosmological observations on cosmic polarization rotation, and (iii) theoretical models of non-minimal interactions of gravitational, electromagnetic and pseudoscalar (axion) fields, and their relevance to cosmology.

## 1. Introduction

One (WTN) of us was a student in Gell-Mann's class of "Topics on Particle Theories" in the late 1960's, and learned ups, downs and the spirit of constructing modern particle theories. Working on thesis in Thorne's group in 1969-1972, he learned relativistic astrophysics and the spirit of phenomenological study of gravitation. With this background, he started to work on the theoretical study and phenomenological study of equivalence principles in the late 1972 in the quiet environment of Bozeman, Montana in Nordtvedt's group. The theoretical work reached *two milestones*, one in 1973 for finding a non-metric theory with pseudoscalar-photon interaction[1] (axion interaction) which observes the Galileo Equivalence Principle and the other in 1974 for proving this is the only non-metric theory in the general χ-g framework of charged particle-electromagnetism system in gravity.[2] The complete paper was written and published[3] after he moved to National Tsing Hua University where he continued to work on both theoretical and phenomenological aspects, and started to work on experimental aspects.[4] In section 2, we review different motivations to reach pseudoscalar interactions and axions together with related developments. In section 3, we present the current status of our precision ellipsometry experiments using high-finesse Fabry-Perot cavity. In the first theoretical *milestone* addressed above, we found that the non-metric pseudoscalar-photon interactions could induce polarization rotation in electromagnetic wave propagation and proposed to use long-distance astrophysical propagation to test the theory.[1] It is fortunate that, due to technological and observational developments, this test has been improved in great precision recently.[4-5] We discuss and compile the recent results in section 4. In the early universe (inflationary or equivalent period, earlier period, etc.), the physics was not yet determined and studies in non-minimal coupling of photons and axions would be warranted. In section 5, we review our recent work on this.[6]

---

[1] Dedicated to Murray Gell-Mann on his 80th birthday.

## 2. Pseudoscalar-photon interaction and axions

In the 5th Patras Workshop on Axions, WIMPs and WISPs held at the University of Durham on 13-17 July 2009, three motivations were presented. In the bottom-up approach,[7] axion is considered as a Goldstone boson associated with spontaneously broken PQ symmetry[8] to fix the strong CP problem. The name of axion was proposed by Wilczek as detergent AXION (AXION is a commercial brand of detergent) to clean up the strong CP problem. As the original axions[9,10] had not been found, invisible axions[11-14] were proposed.

Top-down motivation[15] comes from superstring theory. In supersymmetry/ supergravity, an appropriate action connects gauge and axionic couplings through a single holomorphic function. In type IIA/B superstring theory, axion comes from a Ramond-Ramond antisymmetrical field reduced on the cycle (compactified space). Axions also arise for heterotic string and M-theory. In superstring theory, "the model-independent axion appears in every perturbative string theory, and is closely related to the graviton and dilaton."[16]

The gravity-related motivation[17,18] (historically the first approach as described in the introduction) comes from a phenomenological study of equivalence principles. In 1973, we studied the relationship of Galilio Equivalence Principle (WEP I) and Einstein Equivalence Principle in a framework (the χ-g framework) of electromagnetism and charged particles, and found the following interaction Lagrangian density

$$\mathcal{L}_I = -(1/16\pi)g^{ik}g^{jl}F_{ij}F_{kl} - (1/16\pi)\varphi F_{ij}F_{kl}e^{ijkl} - A_k j^k(-g)^{(1/2)} - \Sigma_I m_I(ds_I)/(dt)\delta(\boldsymbol{x}-\boldsymbol{x}_I), \quad (1)$$

as an example which obeys WEP I, but not EEP.[1-3] ($e^{ijkl}$ is the completely antisymmetric symbol.) The nonmetric part of this theory is

$$\mathcal{L}^{(NM)}_{int} = -(1/16\pi)(-g)^{1/2}\varphi\varepsilon^{ijkl}F_{ij}F_{kl} = -(1/4\pi)(-g)^{1/2}\varphi_{,i}\varepsilon^{ijkl}A_j A_{k,l} \text{ (mod div)}, \quad (2)$$

where 'mod div' means that the two Lagrangian densities are related by partial integration in the action integral. ($\varepsilon^{ijkl} \equiv (-g)^{-1/2}e^{ijkl}$.) The modified Maxwell equations[1-3] are

$$F^{ik}{}_{/k} + \varepsilon^{ikml}F_{km}\varphi_{,l} = -4\pi j^i, \quad (3)$$

where the derivation | is with respect to the Christoffel connection of $g^{ik}$. The Lorentz force law is the same as in metric theories of gravity or general relativity. Gauge invariance and charge conservation are guaranteed. The Maxwell equations (3) are also conformally invariant. Axial-photon interaction induces energy level shift in atoms and molecules, and polarization rotations in electromagnetic wave propagation. Empirical tests of the pseudoscalar-photon interaction (2) were analyzed in 1973; at that time it was only loosely constrained.[1] Now we have effective constraints on polarization rotation in the electromagnetic wave propagation from astrophysical polarization observations and CMB polarization observations for massless or nearly massless case.[4,5,18] Axion with mass is a viable candidate for dark matter search. Recently laboratory experiments have started to give constraints on them.[17]

The rightest term in equation (2) is reminiscent of Chern-Simons[19] term $e^{\alpha\beta\gamma} A_\alpha F_{\beta\gamma}$. There are two differences: (i) Chern-Simons term is in 3 dimensional space; (ii) Chern-Simons term in the integral is a total divergence. A term similar to the one in equation (2) (axion-gluon interaction) occurs in QCD in an effort to solve the strong CP problem with the electromagnetic field replaced by gluon field.[8-10] Carroll, Field and Jackiw[20] proposed a modification of electrodynamics with an additional $e^{ijkl} V_i A_j F_{kl}$ term with $V_i$ a constant vector. This term is a special case of the term $e^{ijkl} \varphi F_{ij} F_{kl}$ (mod div) with $\varphi_{,i} = -\frac{1}{2} V_i$. *Various terms in the Lagrangians discussed here are listed in Table 1 for comparison.*

**Table 1.** Various terms in the Lagrangian and their meaning.

| Term | Dimension | Reference | Meaning |
|---|---|---|---|
| $e^{\alpha\beta\gamma} A_\alpha F_{\beta\gamma}$ | 3 | Chern-Simons[19] (1974) | Intergrand for topological invariant |
| $e^{ijkl} \varphi F_{ij} F_{kl}$ | 4 | Ni[1-3] (1973, 1974, 1977) | Pseudoscalar-photon coupling |
| $e^{ijkl} \varphi F^{QCD}_{ij} F^{QCD}_{kl}$ | 4 | Peccei-Quinn[8] (1977) Weinberg[9] (1978) Wilczek[10] (1978) | Pseudoscalar-gluon coupling |
| $e^{ijkl} V_i A_j F_{kl}$ | 4 | Carroll-Field-Jackiw[19] (1990) | External constant vector coupling |

In the Peccei-Quinn-Weinberg-Wilczek models, axion-photon interaction may or may not be induced. In terms of Feynman diagram, the interaction (2) gives a two-photon-pseudo-scalar vertex and vacuum becomes birefringent and dichroic.[21-25] These effects are quantum in origin, while the cosmic polarization rotation effect discussed following (3) is classical.

Dichroic materials have the property that their absorption constant varies with polarization. For axion models with (2), photon interacts with magnetic field has a cross section to be converted into axion and leaks away. The vacuum with magnetic field becomes absorptive. Since the cross section depends on polarization, so is the absorption. When polarized light goes through vacuum with magnetic field, its polarization is rotated due to difference in absorption in two principal directions of the vacuum for the two polarization components. The polarization rotation ε of the photon beam for light entering the magnetic-field region polarized at an angle of $\theta$ to the magnetic field is

$$\varepsilon = (B^2\omega^2 M^{-2} m_\varphi^{-4}) \sin^2(m_\varphi^2 L/4\omega) \sin(2\theta) \approx B^2 L^2/(16M^2) \sin(2\theta), \quad (4)$$

where $m_\varphi$ is mass of the axion, $M$ the axion-photon interaction energy scale, $\omega$ photon circular frequency and $L$ the magnetic-region length. The approximation is valid in the limit

$$m_\varphi^2 L/4\omega \ll 1. \quad (5)$$

Since axions do not reflect at the mirrors, for multi-passes, the rotation effect increases by number $N$ of passes. Therefore for the case condition (5) is satisfied, the polarization rotation effect is proportional to $NB^2L^2$. Axion leaking away has a cross section to interact with another magnetic field to regenerate photon. Current optical experiments to measure the dichroism and to detect photon regeneration are listed in Table 2 of Ref. 26. In the following section we discuss the method of using high finesse cavity ellipsometry to measure the dichroism and report on the current status of our Q & A experiment.

## 3. High finesse cavity ellipsometry and Q & A experiment

The standard technique to measure birefringence and dichroism is ellipsometry. To measure minute birefringence and minute dichroism, a high finesse cavity is used for enhancing the physical effects. In 1994, two groups – PVLAS (Polarizzazione del Vuoto con LASer) and Q & A (QED [Quantum Electrodynamics] and Axion experiment) – started to build apparatuses using high finesse Fabry-Perot cavity to measure QED birefringence and search for pseudoscalar-photon interactions. PVLAS adapted the earlier scheme proposed in 1979,[27] and had experiences from the participation of some of their members in the BFRT (Brookhaven-Fermilab-Rochester-Trieste) experiment[28] which had used multipass to enhance the physical effects. PVLAS group started to build a vertical Fabry-Perot cavity to accommodate a rotating cryogenic superconducting dipole magnet. We started to build a 3.5m/7m prototype Fabry-Perot cavity with a horizontally rotating permanent dipole magnet for measuring vacuum birefringence and improving the sensitivity of axion search as part of our continuing effort in precision interferometry. In close contact with ground gravitational-wave detection community, we use a number of techniques developed by the community.[29] In June 1994, we met the PVLAS people in the Marcel Grossmann Meeting at Stanford, exchanged a few ideas and encouraged each other. BMV (Biréfringence Magnétique du Vide) group started to construct their experiment[30] using high magnetic field pulses in 2000. Both PVLAS group[31] and Q & A group[32] have used their apparatuses to measure the Cotton-Mouton birefringence of various dilute gases as applications to chemical physics and as calibrations of their apparatuses. The results of two groups in the common cases agree with each other within 1.2 σ. BMV group has also done this recently.[33]

The 2006 report of PVLAS group on the positive detection of dichroism[34] stirred up a lot of experimental activities on optical detection of axions, minicharged particles and related interactions. Among groups working on LSW (Light Shining through the Wall [photon regeneration]) experiments, OSQAR (Optical Search of QED vacuum magnetic birefringence, Axions and photon Regeneration) collaboration also started to build high-finesse ellipsometry for birefringence and dichroism measurement.[35] The original PVLAS results were soon found disfavored by the results of Q & A experiment,[36] and ruled out by further and more careful measurements of PVLAS.[37] Now all groups are working on measuring QED vacuum birefringence as their immediate goal. After this is achieved, the sensitivity for searching axions and other relevant particles would be improved by several orders of magnitude.

We now report on our Q & A experiment. The schematic of the setup of our second phase (2002-2008) is shown in Fig. 1 of ref. 26. The 3.5 m prototype interferometer is formed using a high-finesse Fabry–Perot interferometer together with a high-precision ellipsometer. The two high-reflectivity mirrors of the 3.5 m prototype interferometer are suspended separately from two X-pendulum–double pendulum suspensions mounted on two isolated tables fixed to ground using bellows inside two vacuum chambers. The sub-systems are described in ref. 36. Our results in this phase give $(-0.2 \pm 2.8) \times 10^{-13}$ rad/pass with 18,700 passes through a 2.3 T 0.6 m long magnet for vacuum dichroism measurement, and limit pseudo-scalar-photon interaction and millicharged fermions meaningfully.[36]

We are currently upgrading our interferometer from 3.5 m armlength to 7 m armlength in the 3rd phase. We have installed a new 1.8 m 2.3 T permanent magnet capable of rotation up to 13 cycles per second to enhance the physical effects. We are working with 532 nm Nd:YAG laser as light source with cavity finesse around 100,000, and aim at 10 nrad/Hz$^{1/2}$ optical sensitivity. *With all these achieved and the upgrading of vacuum, vacuum dichroism measurement would be improved in precision by 3-4 orders of magnitude, and QED birefringence would be measured to 28 % in about 50 days.* To enhance the physical effects further, another 1.8 m magnet will be added in the future.

## 4. Constraints on cosmic polarization rotation from observations

For the modified Maxwell equations (3), electromagnetic propagation induces a polarization rotation of angle $\alpha = \Delta\varphi = \varphi_2 - \varphi_1$ where $\varphi_1$ and $\varphi_2$ are the values of the scalar field at the beginning and end of the wave.[1] When the propagation distance is over a large part of our observed universe, we call this phenomenon cosmic polarization rotation.[4,5]

In the CMB (Cosmic Microwave Background) observations, there are variations and fluctuations. The variations and fluctuations due to scalar-modified propagation can be expressed as $\delta\varphi(2) - \delta\varphi(1)$, where 1 denotes a point at the last scattering surface in the decoupling epoch and 2 observation point. $\delta\varphi(2)$ is the variation/fluctuation at the last scattering surface. $\delta\varphi(1)$ at the present observation point is zero or fixed. Therefore the covariance of fluctuation $\langle[\delta\varphi(2) - \delta\varphi(1)]^2\rangle$ gives the covariance of $\delta\varphi^2(2)$ at the last scattering surface. Since our Universe is isotropic to $\sim 10^{-5}$, this covariance is $\sim (\xi \times 10^{-5})^2$ where the parameter $\xi$ depends on various cosmological models.[5,38] Electromagnetic propagation from different directions may acquire different polarization rotations depending on the cosmological structure of the gradient of $\varphi$. If we assume the constant part of the gradient of $\varphi$ dominates, the constraints[4] of CMB observations on the cosmic polarization rotation α from various analyses are updated in the following Table 2.

Table 2. Constraints on cosmic polarization rotation from CMB (cosmic microwave background).

| Reference | Constraint [mrad] | Source data |
|---|---|---|
| Ni[39,40] | ± 100 | WMAP1[41] |
| Feng, Li, Xia, Chen, and Zhang[42] | -105 ± 70 | WMAP3[43] & BOOMERANG (B03)[44] |
| Liu, Lee, Ng[45] | ± 24 | BOOMERANG (B03)[44] |
| Kostelecky and Mews[46] | 209 ± 122 | BOOMERANG (B03)[44] |
| Cabella, Natoli and Silk[47] | -43 ± 52 | WMAP3[43] |
| Xia, Li, Wang, and Zhang[48] | -108 ± 67 | WMAP3[43] & BOOMERANG (B03)[44] |
| Komatsu, *et al.*[49] | -30 ± 37 | WMAP5[49] |
| Xia, Li, Zhao, and Zhang[50] | -45 ± 33 | WMAP5[49] & BOOMERANG (B03)[44] |
| Kostelecky and Mews[51] | 40 ± 94 | WMAP5[49] |
| Kahniashvili, Durrer, and Maravin[52] | ± 44 | WMAP5[49] |
| Wu[53] | 9.6 ± 14.3 ± 8.7 | QuaD (2$^{nd}$ and 3$^{rd}$ sessions)[54] |
| Brown *et al.*[55] | 11.2 ± 8.7 ± 8.7 | QuaD[55] |
| Komatsu *et al.*[56] | -19 ± 22 ± 26 | WMAP7[56] |

Constraints from observations on individual polarization sources are discussed in ref's 4 and 5. The most recent results are the ultraviolet polarization observations of distant radio galaxies.[57] No polarization rotation is detected within a few degrees for each galaxy and overall fitting for a constant scalar gradient/constant vector direction is comparable to the best CMB constraint. More works in this direction are important as they could detect/constrain directional dependence and distinguish cosmological models.

In our original pseudoscalar model, the natural coupling strength $\varphi$ is of order 1. However, the isotropy of our observable universe to $10^{-5}$ may leads to a change of $\Delta\varphi$ over cosmological distance scale $10^{-5}$ smaller. Hence, observations to test and measure $\Delta\varphi$ to $10^{-6}$ is very significant. A positive result may indicate that our patch of inflationary universe has a "spontaneous polarization" in the fundamental law of electromagnetic propagation influenced by neighboring patches and by a determination of this fundamental physical law we could 'observe' our neighboring patches.

The Planck Surveyor was launched in May, 2009.[58] Better sensitivity to $\Delta\varphi$ of $10^{-2}$-$10^{-3}$ (1-10 mrad) is expected. A dedicated future experiment on cosmic microwave background polarization[59-61] may reach $10^{-5}$-$10^{-6}$ $\Delta\varphi$-sensitivity.[39] Astrophysical observations of cosmologically distant objects in various directions will give $\Delta\varphi$ in various directions and will compliment the CMB polarization measurement. Future observations to test and measure $\Delta\varphi$ to $10^{-6}$ and to give $\Delta\varphi$ in various directions are promising.

## 5. Non-minimal coupling of gravitational, electromagnetic and axion fields

To complete the axion interaction theory (1) as a gravitational theory, we have to add a gravitational Lagrangian. This is illustrated by equations (28)-(30) in ref. 4. In the early universe, although inflationary scenario gives the right structure formation, the inflationary physics was not clear and non-minimal extensions of the coupling of photons and axions is worth study. We have formulated a ten-parameter non-minimal extension.[6] The ten non-minimal terms in the Lagrangian together with their physical meaning are compiled in Table 3. $R^{ikmn}$, $R^{mn}$, and $R$ are Riemann tensor, Ricci tensor and Ricci scalar of $g^{mn}$ respectively. $F^*_{mn}$ is the dual tensor of $F^{kl}$. Ref. 6 gives a complete account of these terms together with derivations and some exact solutions. Empirical constraints on coupling parameters from astrophysical birefringence and polarization rotation observations have also been studied. We are currently pursuing on this line further.

Table 3. Ten non-minimal (NM) coupling parameters are divided into four subgroups: the first $q_1$, $q_2$, $q_3$; second $Q_1$, $Q_2$, $Q_3$; third $\eta_1$, $\eta_2$, $\eta_3$; and fourth $\eta_{(A)}$. In the second column the terms in the Lagrangian are given in front of which the corresponding coupling parameters are introduced; the parameters of the first subgroup introduce the terms without the pseudoscalar field $\varphi$; the parameters of the second subgroup relate to the terms linear in $\varphi$; the terms indicated by $\eta_1$, $\eta_2$, $\eta_3$, are quadratic in the four-gradient of $\varphi$; and finally, $\eta_{(A)}$ introduces the term quadratic in $\varphi$. In the last column, we point out the physical meaning of these non-minimal terms, based on decompositions of the constitutive tensors for the electromagnetic (EM) and pseudoscalar fields.

| | Term in the Lagrangian | Physical meaning |
|---|---|---|
| $q_1$ | $(1/2)RF^{mn}F_{mn}$ | NMEM susceptibility linear in the Ricci scalar |
| $q_2$ | $R^{mn}F_{mk}F_n{}^k$ | NMEM susceptibility linear in the Ricci tensor |
| $q_3$ | $(1/2)R^{ikmn}F_{ik}F_{mn}$ | NMEM susceptibility linear in the Riemann tensor |
| $Q_1$ | $(1/2)\varphi RF^{mn}F^*_{mn}$ | NMEM susceptibility induced by the axion with the Ricci scalar |
| $Q_2$ | $\varphi R^{mn}F_{mk}F^{*k}{}_n$ | NMEM susceptibility induced by the axion with the Ricci tensor |
| $Q_3$ | $(1/2)\varphi R^{ikmn}F_{ik}F^*_{mn}$ | NMEM susceptibility induced by the axion with the Riemann tensor |
| $\eta_1$ | $-R^n{}_l F^{ml}\nabla_m\varphi\nabla_n\varphi$ | NMEM current induced by the axion field gradient |
| $\eta_2$ | $-R\nabla^m\varphi\nabla_m\varphi$ | NM axion-graviton derivative coupling with the Ricci scalar |
| $\eta_3$ | $-R^{mn}\nabla_m\varphi\nabla_n\varphi$ | NM axion-graviton derivative coupling with the Ricci tensor |
| $\eta_{(A)}$ | $R\varphi^2$ | NM correction to the mass square of the axion |

We are grateful to the National Science Council (NSC 98-2112-M-007-009) for support.